\documentclass[12pt]{article}

\begin{document}

\title{Form Factor of a Quantum Graph \\ in a Weak Magnetic Field}

\author{Taro Nagao and Keiji Saito$^{\dagger}$}
\date{}
\maketitle

\begin{center}
\it
Institut f\"ur Theoretische Physik, Universit\"at zu K\"oln, \\ 
Z\"ulpicher Str. 77, 50937 K\"oln, Germany \\  
(Permanent Address: Department of Physics, Graduate 
School of Science, \\ Osaka University, Toyonaka, Osaka 
560-0043, Japan) 
\end{center}
\begin{center}
\it
$^{\dagger}$Department of Applied Physics, Graduate School of 
Engineering, University of Tokyo, Bunkyo-ku, Tokyo 113-8656, 
Japan
\end{center}

\bigskip
\begin{center}
\bf Abstract 
\end{center}
\par
\bigskip
\noindent
Using periodic orbit theory, we evaluate the form factor 
of a quantum graph to which a very weak magnetic field 
is applied. The first correction to the diagonal approximation 
describing the transition between the universality classes is 
shown to be in agreement with Pandey and Mehta's formula of 
parametric random matrix theory. 

 
\newpage
\noindent
\section{Introduction}
\setcounter{equation}{0}
\renewcommand{\theequation}{1.\arabic{equation}}
Since more than a decade, it has been known 
that the energy level statistics of classically 
chaotic systems universally follows the prediction of random 
matrix theory. As the semiclassical theory of chaotic systems 
is described in terms of classical periodic orbits, 
it is preferable if we can also understand the universal 
level statistics in terms of them. However, until recently, 
such an understanding had been limited within the framework 
of Berry's diagonal approximation for double sums over 
periodic orbits\cite{BERRY}. 
\par
A recent breakthrough was brought about by 
Sieber and Richter\cite{SR01,RS02}. They identified the pairs 
of periodic orbits giving the first off-diagonal 
correction and showed that such pairs indeed 
exist in chaotic systems. The first 
off-diagonal correction was then found to be in 
agreement with random matrix theory. 
Berkolaiko, Schanz and Whitney rederived 
the same correction term for 
quantum graphs\cite{BSW02} and further succeeded in 
evaluating the second correction\cite{BSW02-2}. 
\par
In this paper, we investigate a quantum graph and 
extend the calculation of the first off-diagonal 
correction to the case intermediate between 
the absence and presence of a magnetic field. 
In random matrix theory, this intermediate case corresponds 
to Pandey and Mehta's two matrix model\cite{PM,MP}. Chaotic systems 
with time reversal symmetry (without magnetic field) are 
described by GOE(Gaussian Orthogonal Ensemble) of 
random matrices. On the other hand, when time reversal 
symmetry is broken by the application of a magnetic field, 
GUE(Gaussian Unitary Ensemble) becomes suitable. 
Pandey and Mehta formulated an intermediate random matrix 
ensemble (two matrix model) between GOE and GUE and were 
able to derive the correlation functions among the eigenvalues. 
Within the diagonal approximation, Bohigas et al.\cite{BGOAS} already showed 
that the periodic orbit theory was in accordance with the 
intermediate random matrix ensemble. We will show that, by applying 
Berkolaiko, Schanz and Whitney's method, the first off-diagonal 
correction is also in agreement with Pandey and Mehta's formula. 
\par   
Let us explain the model. We consider quantum mechanics on a graph
\cite{KS,SS1,SS2,BK}. 
A graph consists of vertices connected by bonds. For example, 
in a globally coupled graph, every vertex is connected to all 
vertices, while in a star graph, one particular central 
vertex is connected to all of the others. 
\par
On the bonds a particle behaves like a free particle 
and on the vertices it is scattered according to a given scattering 
matrix. On the bond $(j,l)$ connecting the $j$-th 
vertex and $l$-th vertex, the Schro\"odinger equation   
\begin{eqnarray}
\left( -i {{\rm d} \over {\rm d}x_{jl}} - A_{jl} \right) \Psi (x_{jl}) 
&=& k^2 \Psi (x_{jl})
\label{seq}
\end{eqnarray}
holds, where $A_{jl}$ is a magnetic vector potential satisfying  
$A_{jl} = - A_{lj}$. Therefore, on the bond $(j,l)$, 
the wave function $\Psi (x_{jl})$ is proportional to
\begin{eqnarray}
\exp \left( i k x_{jl} + i A_{jl}  x_{jl} \right). 
\end{eqnarray}
The relative amplitude of this wave function is decided by the 
scattering matrices on the vertices. Let us suppose that, 
if a wave function has the amplitude $1$ on 
the bond $(j,l)$, then the amplitude of the wave function on the 
bond $(l,m)$ is $\sigma_{j,m}^{(l)}$. Then the scattering matrix connecting 
the bond $(j,l')$ and $(l,m)$ is 
\begin{eqnarray}
S_{j l' , l m} &=& \sigma_{j m}^{(l)} 
\exp \left( i k L_{l m} + i A_{l m}  L_{l m} \right)
\delta_{l, l'},
\end{eqnarray}
where $L_{jl}$ is the length of the bond $(j,l)$ 
satisfying $L_{jl} = L_{lj}$. The total number of 
the vertices and directed bonds are denoted by 
$N$ and $B$, respectively. In this paper, we consider only a 
globally coupled graph (every vertex is connected to all vertices), 
so that $B$ is equal to $N^2$. We further assume that the scattering 
matrix is given by the discrete Fourier transform
\begin{eqnarray}
\label{SIGMA}
\sigma_{m n }^{(l)} &=& \frac{1}{\sqrt{N}} e^{2\pi i mn /N}.
\end{eqnarray}
\par
Chaotic dynamics in the classical limit is 
characterized by an approach to an equidistribution 
over all bonds. Noting that an analogue of the classical 
Frobenius-Perron operator is given by $M_{m' l,l m} 
\equiv |S_{m' l,l m}|^2$, we can write it as 
\begin{equation}
\label{EQUI} 
\lim_{t \rightarrow \infty} M_{m' l,l m}^t = 1/B.  
\end{equation}
For the scattering matrix (\ref{SIGMA}), (\ref{EQUI}) 
holds even for finite $t$. Berkolaiko, Schanz and Whitney\cite{BSW02} 
argued that, if the convergence of (\ref{EQUI}) in the limit 
$t \rightarrow \infty$ is sufficiently fast, the 
form factor is in agreement with random matrix theory. 
We expect that a similar condition holds in the intermediate 
case. However it is not discussed here and left for 
future works.   

\section{Periodic Orbit Theory}
\setcounter{equation}{0}
\renewcommand{\theequation}{2.\arabic{equation}}
Let us define $P$ and $Q$ as sequences of 
vertices $[p_1,p_2,\cdots,p_t]$ and $[q_1,q_2,\cdots,q_t]$, 
respectively. Each of $p_j$ and $q_j$ takes the values 
$1,2,\cdots,N$. According to Berkolaiko, Schanz and 
Whitney\cite{BSW02}, the form factor $K(\tau)$ can be written 
in terms of the scattering matrices as 
\begin{eqnarray}
K (\tau ) &=& \frac{1}{B} \langle  | {\rm Tr }   S^{t} |^2 \rangle \nonumber \\
&=&
\frac{1}{B} \lim_{\kappa \to \infty } 
\kappa^{-1} \sum_{P,Q} \int_{0}^{\kappa}
d k \, 
\sigma_{p_t p_2}^{(p_1)} {\rm e}^{i L_{p_1 p_2} (k + A_{p_1 p_2})}
\sigma_{p_1 p_3}^{(p_2)} {\rm e}^{i L_{p_2 p_3} (k + A_{p_2 p_3})} \cdots \nonumber \\
&&\cdots
\sigma_{p_{t-1} p_1}^{(p_t)} {\rm e}^{i L_{p_t p_1} (k + A_{p_t p_1})}
{\sigma_{q_t q_2}^{(q_1)}}^{\ast} {\rm e}^{-i L_{q_1 q_2} (k + A_{q_1 q_2})} 
\cdots 
{\sigma_{q_{t-1} q_1}^{(q_t)}}^{\ast} {\rm e}^{-i L_{q_t q_1} (k + A_{q_t q_1})} 
\nonumber \\
&=& 
\frac{1}{B} \sum_{P,Q} {\cal A}_{P} {{\cal A}_{Q}}^{\ast}
{\rm e}^{i\left( 
L_{p_1 p_2} A_{p_1 p_2} + \cdots + L_{p_t p_1} A_{p_t p_1} - 
L_{q_1 q_2} A_{q_1 q_2} - \cdots - L_{q_t q_1} A_{q_t q_1}
\right)} \delta_{L_P,L_Q}, \nonumber \\ 
\end{eqnarray}
where   
\begin{eqnarray}
 {\cal A}_{P} & = & \sigma_{p_t p_2}^{(p_1)} \cdots 
              \sigma_{p_{t-1} p_1}^{(p_t)}, 
 \ \ \ L_P = L_{p_1 p_2} + \cdots + L_{p_t p_1},   
     \\
 {\cal A}_{Q} & = & \sigma_{q_t q_2}^{(q_1)} \cdots 
              \sigma_{q_{t-1} q_1}^{(q_t)},   
 \ \ \ L_Q = L_{q_1 q_2} + \cdots + L_{q_t q_1}   
\end{eqnarray}
and
\begin{equation}
\tau = t/B.
\end{equation}
We are interested in the scaling limit $t \rightarrow \infty$, 
$B \rightarrow \infty$ with $\tau$ fixed. Then we expect 
that the diagonal term and its corrections give the 
expansion around $\tau = 0$. 
\par
The diagonal term comes from the cases 
in which circular permutations of $P$ and $Q$ 
(or the reversal of $Q$) coincide. 
It is explicitly written as 
\begin{eqnarray}
K^{ \rm diag } (\tau )
&=&{t \over B} \sum_{P} |{\cal A}_{P} |^2 \left[
e^{2i\sum_{j} L_{p_{j} p_{j+1}} A_{p_{j} p_{j+1}}  }
+ 1
\right], 
\end{eqnarray}
where $p_{t+1} \equiv p_1$. On the other hand, 
the first off-diagonal correction  $K^{\rm off}(\tau)$ 
comes from the pairs of self-intersecting orbits 
differing in the orientation of a single loop. 
We suppose that the orbits are self-intersecting 
at a vertex $\alpha \equiv p_1 = p_{t'}$. 
Then, in the first off-diagonal correction, 
the orbit sum is taken over the pair 
$P=[\alpha,l_1,\alpha,l_2]$ and 
$Q=[\alpha,l_1,\alpha,\bar{l}_2]$. Here 
$l_1=[p_2,p_3,\cdots,p_{t'-1}]$, 
$l_2=[p_{t'+1}, p_{t'+2},\cdots,p_t]$, 
respectively, and ${\bar l}_2 = [p_t,p_{t-1},\cdots,
p_{t'+1}]$ is the reversal of the 
sequence $l_2$. That is, $K^{\rm off}(\tau)$ can be 
written as 
\begin{eqnarray}
\label{KOFF}
K^{\rm off}(\tau) &=& 
{t^2 \over B} \sum_{t'=4}^{t-2} {\sum_{P}}^{\prime} 
\left( 1- \delta_{cd} \right)
|\sigma_{\alpha p_3}^{(a)}|^2 \cdots
|\sigma_{p_{t'-2} \alpha}^{(b)}|^2  
|\sigma_{\alpha p_{t' +2}}^{(c)}|^2 \cdots
|\sigma_{p_{t-1} \alpha}^{(d)}|^2 \nonumber \\
&& \times 
\sigma_{d a}^{(\alpha)}  \sigma_{b c}^{(\alpha)}  
{\sigma_{c a}^{(\alpha)}}^{\ast}  
{\sigma_{b d}^{(\alpha)}}^{\ast}  
\exp\left[ 2i \left( L_{d\alpha} A_{d\alpha} + 
L_{p_{t-1} p_t} A_{p_{t-1} p_t}  + \cdots +
L_{p_{t'} p_{t' +1}} A_{p_{t'} p_{t' +1}} 
\right)\right]. \nonumber \\ 
\end{eqnarray}
Here $a,b,c,d$ are identified with 
$p_2,p_{t'-1},p_{t'+1}, p_t$, respectively. 
As indicated by the prime, 
the sum over $P$ obeys 
the restriction that $l_1$ is not 
identical to the reversal of 
itself($l_1 \neq {\bar l}_1$). 
\par
We are now in a position to calculate the diagonal 
contribution and the first off-diagonal correction. 
For simplicity, we set $L_{jl} = 1/2$ from now on. 
\par
\medskip
\noindent
(1) the diagonal contribution 
\par
\medskip
\noindent
\begin{eqnarray}
K^{ \rm diag } (\tau )
&=&{t \over B} \sum_{P} |{\cal A}_{P} |^2 \left[
e^{2i\sum_{j} L_{p_{j} p_{j+1}} A_{p_{j} p_{j+1}}  }
+ 1
\right] \nonumber \\
&=& 
{t \over B} \sum_{P} {1\over N^{t} } \left[
e^{2i\sum_{j} L_{p_{j} p_{j+1}} A_{p_{j} p_{j+1}}  }
+ 1
\right] \nonumber \\
&=& 
{t \over B} {1\over N^{t} } \left[
N^t + {\rm Tr} g^t \right].
\end{eqnarray}
Here the elements $g_{jl}$ of an $N \times N$ 
transfer matrix $g$ is defined as
\begin{equation}
g_{jl} = \left\{ \begin{array}{ll} 1, & j=l, \\ 
{\rm e}^{i A_{jl}}, & j \neq l. \end{array} \right.
\end{equation}
Since we are interested in the long time limit $t \rightarrow \infty$, 
we only need to know the behavior of the largest eigenvalue $\Lambda$ 
of $g$. For small $A_{jl}$, the largest eigenvalue $\Lambda$ can be 
evaluated by the perturbation method. Let us decompose $g$ as   
\begin{equation}
g=g_0 + g',
\end{equation}
where $g_0$ is an $N \times N$ matrix with all the elements equal 
to $1$. Then the largest eigenvalue of $g_0$ is $N$ and the corresponding 
normalized eigenvector ${\bf x}_1$ is $(1/\sqrt{N},1/\sqrt{N},
\cdots,1/\sqrt{N})^{T}$ (Here a superscript $T$ means a transpose). 
All the other eigenvalues of $g_0$ are $0$ and let us denote the corresponding 
orthonormalized eigenvectors by ${\bf x}_2,{\bf x}_3,\cdots,{\bf x}_N$. 
Then we regard $g'$ as a small perturbation and apply the 
standard technique of perturbation theory to evaluate the 
largest eigenvalue. The second order perturbation gives  
\begin{equation}
\Lambda \sim N + {\bf x}_1^{T} g' {\bf x}_1 
+ \frac{1}{N} \sum_{n=2}^N |{\bf x}_n^{T} g' 
{\bf x}_1 |^2,
\end{equation}
which yields
\begin{equation}
\Lambda \sim N -\frac{1}{N} \sum_{j<l} A_{jl}^2 + 
\frac{1}{N} \sum_{n=2}^N |{\bf x}_n^{T} a 
{\bf x}_1 |^2,
\end{equation}
where $a$ is an $N \times N$ matrix with the elements 
\begin{equation}
a_{jl} = \left\{ \begin{array}{ll} 0, & j=l, \\
 i A_{jl}, & j \neq l. \end{array} \right.
\end{equation}
\par
In the limit $N \rightarrow \infty$, the sum $C$ of the contributions 
from the second and third terms is normally $O(N)$ (for fixed $A_{jl}$). 
We then generally write $C \sim - N b$ with $b$ proportional to the 
magnetic field applied to the graph. For example, if $\sum_{l \neq j} A_{jl} 
= 0$ for all $j$ and $|A_{jl}| = A$ for all $j,l$ (this is possible when 
$N$ is odd), the third term vanishes and the second term gives 
$C = -(N-1)A^2/2 \sim -N A^2/2$, so that $b = A^2/2$. The trace 
of the multiples of the matrix $g$ can be estimated as
\begin{equation}
\label{TRG}
{\rm Tr}g^t \sim \Lambda^t \sim (N - N b)^t \sim N^t {\rm e}^{- bt}
\end{equation}
with $t \rightarrow \infty$, $b \rightarrow 0$ and $bt$ fixed. Putting this, 
we arrive at 
\begin{equation}
K^{\rm diag } (\tau ) \sim {t \over B} (1 + {\rm e}^{-bt}).
\end{equation}
This is the diagonal contribution to the form factor in a 
weak magnetic field, as we shall see below, in agreement with 
random matrix theory. We now proceed to the first off-diagonal 
correction.   
\par
\medskip
\noindent
(2) the first off-diagonal correction 
\par
\medskip
\noindent
As noted before, the contribution $K^{ \rm off }_{\rm srt} (\tau)$ 
from self-retracing orbits with $l_1 = {\bar l}_1$ is removed from 
the periodic orbit sum (\ref{KOFF}) for $K^{\rm off}(\tau)$, since it is 
already included in the diagonal contribution. In spite of that, 
it is convenient to first consider the sum with no such restriction. 
The periodic sum including the self-retracing orbits 
can be readily evaluated as  
\begin{eqnarray}
K^{\rm off } (\tau ) + K^{ \rm off }_{\rm srt} (\tau ) 
&=& 
{t^2 \over B} \sum_{t'=4}^{t-2}
N^{t'-t-2}
\sum_{\alpha abcd} \left( 1- \delta_{cd} \right)
g_{\alpha c} \left[ g^{t-t'-1}\right]_{c,d} g_{d \alpha}
\sigma_{d a}^{(\alpha)}  \sigma_{b c}^{(\alpha)}  
{\sigma_{c a}^{(\alpha)}}^{\ast}  
{\sigma_{b d}^{(\alpha)}}^{\ast}  \nonumber \\
&=& 
{t^2 \over B} \sum_{t'=4}^{t-2}
N^{t'-t-2}
\sum_{\alpha cd} \delta_{cd} \left( 1- \delta_{cd} \right)
g_{\alpha c}\left[ g^{t-t'-1}\right]_{c,d} g_{d \alpha }
 = 0.
\end{eqnarray} 
Here we utilized the unitarity of the matrices 
$\sigma^{(\alpha)}$.  As to the contribution from the 
self-retracing orbits, if we neglect completely 
self-retracing paths which are exponentially few, 
a diagrammatic cancellation leaves only the terms with 
$t' = 4$ and $t' = 5$, in which the factors 
$1 - \delta_{cd}$ are removed\cite{BSW02}. 
Therefore we find  
\begin{eqnarray}
\label{KOFF2}
K^{ \rm off }(\tau ) = - K^{ \rm off }_{\rm srt} (\tau ) 
&=& - {t^2 \over B}  \left[
N^{2-t} \sum_{\alpha acd} 
\sigma_{d a}^{(\alpha)}  \sigma_{a c}^{(\alpha)}  
{\sigma_{c a}^{(\alpha)}}^{\ast}  {\sigma_{a d}^{(\alpha)}}^{\ast}  
g_{\alpha c}\left[ g^{t-5}\right]_{c,d} g_{d \alpha } \right. \nonumber \\ 
& & \left. + N^{3-t} \sum_{\alpha acd} \sigma_{d a}^{(\alpha)}  
\sigma_{a c}^{(\alpha)}  {\sigma_{c a}^{(\alpha)}}^{\ast}  
{\sigma_{a d}^{(\alpha)}}^{\ast}  
g_{\alpha c}\left[ g^{t-6}\right]_{c,d} g_{d \alpha } \right] \nonumber \\
&=& -  
{t^2 \over B} \left[ N^{1-t} {\rm Tr} \left( g^{t-3}\right) + 
N^{2-t} {\rm Tr} \left( g^{t-4}\right)
\right].
\end{eqnarray}
Substitution of (\ref{TRG}) into (\ref{KOFF2}) leads to 
\begin{equation}
K^{\rm off }(\tau) \sim  - 2 {t^2 \over B^2} {\rm e}^{-bt}.
\end{equation} 
As a result, the diagonal and the first off-diagonal terms 
are summed up to yield    
\begin{equation}
K^{\rm diag}(\tau) + K^{\rm off}(\tau) \sim  \tau + 
\tau {\rm e}^{-bt} - 2 \tau^2 {\rm e}^{-bt}.
\end{equation}.
 
\section{Random Matrix Result: Pandey-Mehta Formula}
\setcounter{equation}{0}
\renewcommand{\theequation}{3.\arabic{equation}}
In this section we calculate the Fourier transform 
of random matrix result (Pandey-Mehta formula) to derive a prediction of 
the form factor $K(\tau)$. Let us define the Fourier transform of 
the two energy level correlation function $Y(r ; \rho )$ as $Y(k; \rho)$. 
The form factor $K(k)$ can be then written as 
\begin{equation}
K( k) = 1- Y(k; \rho).
\end{equation}
Pandey and Mehta's formula \cite{PM} is 
\begin{equation}
Y(r ; \rho ) =\left( {\sin \pi r} \over \pi r\right)^{2} - 
\frac{1}{\pi^2} \int_{0}^{\pi} d k_1 \int_{\pi}^{\infty} d k_2 \,
\left( {k_1 \over k_2} \right) \sin (k_1 r )\sin (k_2 r )
e^{2 \rho^2 (k_1 - k_2)(k_1 + k_2)}.
\end{equation}
Here $r$ is the distance between the two energy levels and  
$\rho$ is a parameter corresponding to a weak magnetic 
field. In the limit $\rho \rightarrow 0$, 
$Y(r;\rho)$ becomes the two level correlation function of 
GOE, while, in the limit $\rho \rightarrow \infty$, 
$Y(r;\rho)$ approaches that of GUE. 
\par   
The Fourier transform $Y(k; \rho)$ can be evaluated as 
\begin{equation}
Y(k; \rho) =  
1 - k - {1\over 2\pi} 
 \int_{\pi - \bar{k}}^{\pi} d k_1 {k_1 \over k_1 + \bar{k}}
e^{-2 \rho^2 ( 2 k_1 + \bar{k} ) \bar{k} }, \ \ \  0 \le {\bar k} \le \pi, 
\end{equation}
where $\bar{k}  = 2\pi k$. Let us define
\begin{equation}
F({\bar k}) =  {1\over 2\pi} 
 \int_{\pi - \bar{k}}^{\pi} d k_1 {k_1 \over k_1 + \bar{k}}
e^{-2 c ( 2 k_1 + \bar{k} )}
\end{equation}
with $c = \rho^2 {\bar k}$ fixed. Then, for small ${\bar k}$, 
$F({\bar k})$ can be expanded as
\begin{equation}
F({\bar k}) = F(0) + F^{\prime}(0) {\bar k} + \frac{1}{2} 
F^{\prime \prime}(0) {\bar k}^2 + \cdots,
\end{equation}
where $F^{\prime}({\bar k})$ and $F^{\prime \prime}({\bar k})$ 
are the first and second derivatives of $F({\bar k})$, 
respectively. The derivatives are readily evaluated as
\begin{eqnarray}
F^{\prime}(0) & = & \frac{1}{2 \pi} {\rm e}^{- 4 \pi c}, \nonumber \\ 
F^{\prime \prime}(0) & = & - \frac{1}{\pi^2} {\rm e}^{- 4 \pi c},  
\end{eqnarray} 
which lead to (for small $k$)
\begin{eqnarray}
Y(k; \rho) & \sim & 1 - k - F^{\prime}(0) {\bar k} - \frac{1}{2} 
F^{\prime \prime}(0) {\bar k}^2  \nonumber \\ 
& = & 1 - k - k {\rm e}^{- 4 \pi c} 
+ 2 k^2 {\rm e}^{- 4 \pi c}.
\end{eqnarray}
Therefore we arrive at  
\begin{equation}
K(k) \sim k + k {\rm e}^{- 4 \pi c} 
- 2 k^2 {\rm e}^{- 4 \pi c},
\end{equation}
which is in agreement with periodic orbit theory 
with an identification
\begin{equation}
\tau = k, \ \ \ bt = 4 \pi c = 8 \pi^2 \rho^2 k.  
\end{equation}
\par
In summary, using the periodic orbit theory we evaluated 
the first off-diagonal correction to the form factor of 
a quantum graph in a very weak magnetic field. It was 
found that the result was consistently in agreement with 
Pandey-Mehta formula of random matrix theory. The extension 
to the second off-diagonal correction seems promising, 
since it was already worked out in the absence of a 
magnetic field \cite{BSW02-2}.

\end{document}